\documentstyle[preprint,aps,epsfig]{revtex}
\tightenlines
\begin{document}
\title{Determination of the proper embedding parameters \\
for noisy time series}

\author{Jeong-No Lee$^{1,2}$\footnote{Electronic mail:~jnlee@phys.kyungwon.ac.kr} and Kwang-Sup Soh$^1$\footnote{Electronic mail:~kssoh@phyb.snu.ac.kr}}
\address{$^1$ Department of Physics Education, Seoul National University, Seoul 151-742, Korea \\
$^2$ Department of Physics, Kyungwon University, Sungnam, Kyunggi-do 461-701, Korea}
\maketitle
\thispagestyle{empty}

\begin{abstract}
We suggest an algorithm for determining the proper delay time and the minimum embedding dimension for Takens' delay-time embedding procedure.
This method resorts to the rate of change of the spatial distribution of points on a reconstructed attractor with respect to the delay time, and can be successfully applied to a noisy time series which is too noisy to be discriminated from a structureless noisy time series by means of the correlation integral, and also indicates that the proper delay time depends on the embedding dimension.
\ \\[5mm] 
\end{abstract}

\pagebreak

Since the discovery of the method to reconstruct state spaces from a time series\cite{packard,takens},
various attempts have been implemented to choose the embedding parameters, the proper delay time and the minimum embedding dimension\cite{schuster,fraser,broomhead,liebert,buzug,lipton,kennel,kaplan}.
By the theorem of Takens\cite{takens}, the reconstructions are generically diffeomorphic to the original dynamics if $d > 2d_{A}$ where $d_{A}$ is the dimension of the compact invariant smooth manifold $A$ corresponding to the continuous time dynamical system and $d$ is the embedding dimension.
From the theorem, one can not, however, determine the delay time and the minimum embedding dimension.
Nevertheless because the quality of the reconstructed attractor affects strongly the success of noise reduction processes\cite{hegger} and forecasting methods as well as accurate calculations of various invariant quantities such as Lyapunov exponents, it is essential to choose the proper embedding parameters before carrying out the above mentioned jobs.

In computational analysis with a noisy time series, because what the proper delay time is depends on the purpose for reconstructing attractors, it is suggested that the delay time is much less critical to success of the reconstruction process than the embedding dimension.

In general, with the chosen delay time too small, the reconstructed points coincide approximately within the range of errors so that the reconstructed attractor appears ``stretched out and randomly crowded'' along the identity line in the time-delayed coordinates. If the delay time is too large, operations such as Grassberger-Procaccia algorithm\cite{grassberger} for the correlation dimension can be lead to wrong results\cite{caputo} due to the nonlinearity of the embedding process not of the original dynamics.

On the other hand, working in any larger dimension than the minimum leads to a more computational burden in obtaining the invariant quantities such as Lyapunov exponents, prediction, etc.
And it also enhances the problem of contamination by round-off or measurement errors since the noises will populate and dominate the additional dimensions of the embedding space.
The most important reason for looking for the minimum embedding dimension is to obtain an intuition for understanding and modeling the underlying dynamics of a noisy time series.
Also in many schemes for noise reduction, the first job to do is to have at least a vague idea about the minimum embedding dimension\cite{hegger}.

In this paper, we propose a method to determine simultaneously the minimum embedding dimension $d_{M}$ and the proper delay time $T_{d}^{d}$ for a noisy time series.
The suggested method can be applied successfully to a noisy time series contaminated by a measurement noise of which the size is so much large that one can not disentangle the deterministic part from the random part by means of the correlation integral\cite{avraham}.
Also on the basis of this method, we find that the proper window length $W^{d} \equiv (d-1)T_{d}^{d}$ remains approximately constant if $d \geq d_{M}$.

Let a continuous signal $w(t)$ be measured at each ``sampling interval'' $T_{s}$ to yield a time series $\{v(k)\}$:
\begin{equation}
\{v(k)\} \equiv \{v(k)~|~v(k) = w(kT_{s}),~~ k=1,2,\cdots,N_{T}\},  \label{time-series}
\end{equation}
where $N_{T}$ is the total number of time series.
We then assign to ${\vec x}(t)$ the delay vector:
\begin{eqnarray}
{\vec y}(p|d,T_{d}) = (v(1+(p-1)j), v(1+(p-1)j+T_{d}), \cdots, v(1+(p-1)j+(d-1)T_{d}))  \label{emb-vec} \\
~~~~~~~~~~~~p=1,2, \cdots,N  \nonumber
\end{eqnarray}
In ${\vec y}(p|d,T_{d})$, $d$ and $T_{d}$ represent the embedding dimension and the delay time in the unit of the sampling time $T_{s}$, respectively.
$j$ is the interval in the unit of sampling time $T_{s}$ between the first components of successive delay vectors, and describes how the time series is sampled to create a set of delay vectors with a computationally adequate size from a given time series.
The time $(d-1)T_{d}$, spanned by each delay vector, is called the ``window length'' of the embedding\cite{broomhead,albano}.
In general, the coordinates of the delay vector made in this way are nonlinearly represented by the original dynamical variables.
This is one of the reasons that the singular value decomposition is not appropriate as a method to choose the proper embedding dimension\cite{fraser2,mees}.
Also once the Takens' criterion $d>2d_{A}$ is satisfied and the delay time is not too small or too large, it is difficult to say which delay time is the most proper.
Whether the delay time is the most adequate or not depends on the operation applied to the reconstructed attractor.

In this paper, we propose as the criterion for the proper value of the delay time $T_{d}$ the first minimum of the rate of change of the distribution of points on a reconstructed attractor as a function of the delay time $T_{d}$.
The increase of the delay time $T_{d}$ from enough small a value let the reconstructed points go away from the identity line in the reconstructed coordinates, that is, the reconstructed attractor expands spatially in the reconstructed coordinates.
Due to the boundness of attractors, the rate of expansion monotonically decreases and the folding effect becomes larger as the delay time $T_{d}$ increases.
At last, the folding effect is equal to the expanding one. At this point, the rate of variation of the spatial distribution of attractor becomes the first minimum value $T_{d}^{d}$. This criterion we suggest is almost equivalent to the maximal expansion condition of the reconstructed attractor.

To infer the distribution of points of a attractor, we introduce $D(N,r,d,T_{d})$ which is defined as the fraction of pairs of delay vectors between which the distance lies from $r$ to $r+\Delta r$:
\begin{equation}
D(N,r,d,T_{d}) \equiv  \frac{2}{N(N-1)}\sum^{N-1}_{i=0}\sum^{i-1}_{j=0}H(r), \label{int-quan}
\end{equation}
where 
\begin{equation}
H(r) = \left\{ \begin{array}{ll}
		1 & \mbox{~~~~~if $r \leq \sqrt{|{\vec y}(i|d,T_{d})-{\vec y}(j|d,T_{d})|^{2}/d} < r + \Delta r$} \\
		0 & \mbox{~~~~~otherwise}
	      \end{array} \right.    \label{h(r)}
\end{equation}
with $\Delta r$ small enough.
In Eq.(\ref{h(r)}), $|\cdot|^{2}$ means the square of Euclidean distance between ${\vec y}(i|d,T_{d})$ and ${\vec y}(j|d,T_{d})$ and is devided by the embedding dimension $d$ in order to eliminate the effect that the size of a reconstructed attractor becomes larger in a higher embedding dimension $d$.
Fixing the embedding dimension $d$, at each delay time $T_{d}$ we calculate Eq.(\ref{int-quan}) and the sum of diffenences between the adjacent delay times, $S(N,d,W=(d-1)T_{d})$:
\begin{eqnarray}
S(N,d,W=(d-1)T_{d}) & \equiv & \int_{0}^{R_{A}} |D(N,r,d,T_{d}+\Delta T_{d})-D(N,r,d,T_{d})|dr  \nonumber \\
	& \approx & \Delta r \sum_{k=1}^{R_{A}/\Delta r} |D(N,k\Delta r,d,T_{d}+\Delta T_{d})-D(N,k\Delta r,d,T_{d})|    \label{sum-diff}
\end{eqnarray}
where $R_{A}=\sqrt{R^{2}/d}$ with $R$ the largest length among the distances between any two points, and $\Delta T_{d}$ and $\Delta r$ are small enough.
The first minimum point $T_{d}^{d}$ of $S(N,d,W=(d-1)T_{d})$ corresponds to the proper delay time in the embedding dimension $d$.

\begin{figure}
\epsfig{figure=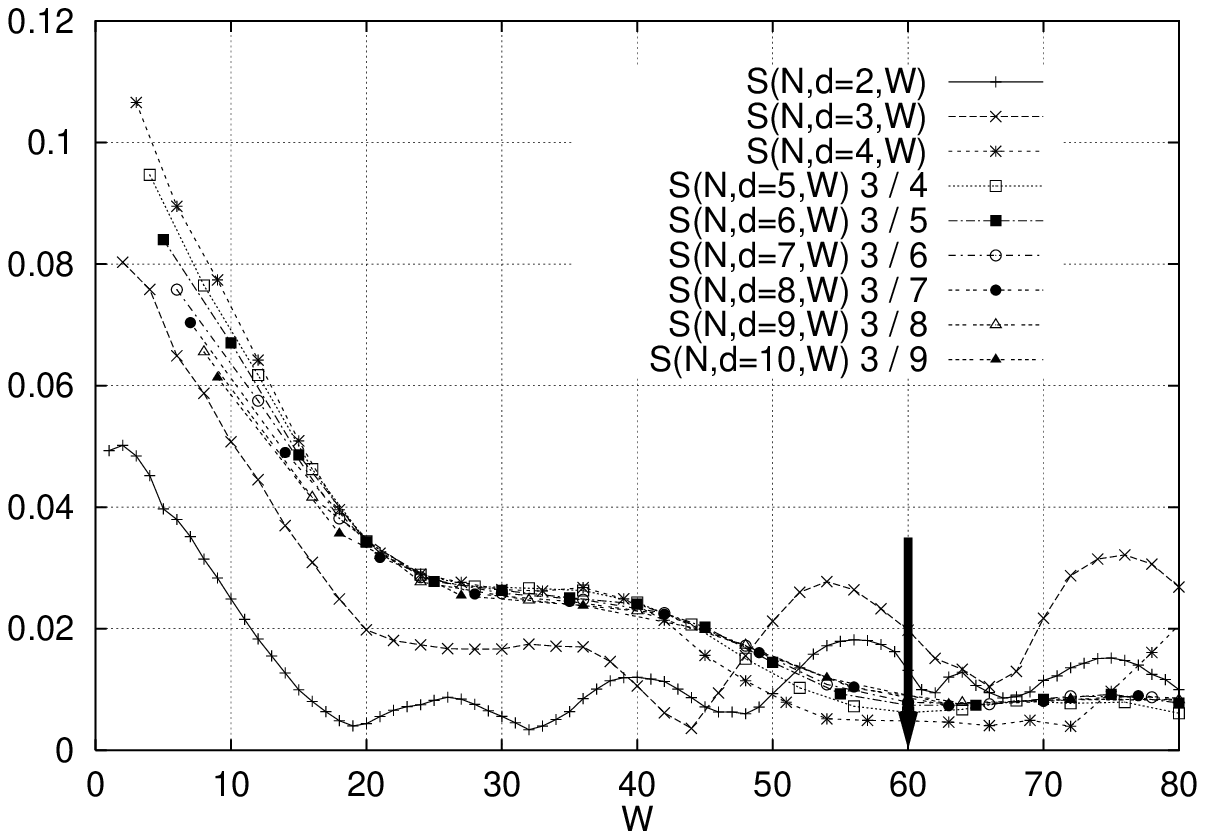}
\caption{$S(N,d,W)$ vs window length $W$ for $N=5000$, evenly picked up among $50000$ points generated by the Hyperchaos system of R\"{o}ssler. For the sake of comparison, each $S(N,d,W)$ was appropriately scaled. The thick arrow points the proper window length $W^{d} \approx 60$ $(d \geq d_{M})$. We can identify the true value of the minimum embedding dimension $d_{M}=4$. Therefore the proper delay time is $W^{d=4}/3 = T_{d}^{d=4} \approx 20$. On the other hand, the proper window lengths in the embedding dimension $d=2$ and $d=3$ are $W^{d=2} \approx 20$ and $W^{d=3} \approx 45$, respectively. $\Delta r = 2$, $\Delta T_{d}=T_{s}=0.1$.}   \label{hyper-nonoise} 
\end{figure}

On the other hand, while reconstructed attractors with an embedding dimension $d$ less than the minimum embedding dimension $d_{M}$ have geometric overlaps,
attractors reconstructed under the condition of $d \ge d_{M}$ do not have any overlap, and are diffeomorphic to each other.
These facts mean that $S(N,d \geq d_{M},W)$s' show a similar behaviour with respect to the window length $W=(d-1)T_{d}$, at least until the first minimum point $W^{d}=(d-1)T_{d}^{d}$ of $S(N,d,W)$.
After the first minimum point $W^{d}=(d-1)T_{d}^{d}$ is passed through, the folding effect of the embedding procedure dominates the expanding effect.
The folding effect raises an unexpected result in the calculation of $D(N,r,d,T_{d})$ because in the range from $r$ to $r+\Delta r$, points even from a folded part are counted. This is just the reason that if the window length is too large, Grassberger-Procaccia algorithm for the correlation dimension can be lead to wrong results\cite{caputo}.
So we can choose simultaneously the proper embedding parameters through the calculation of $S(N,d,W)$s'.
Fig.\ref{hyper-nonoise} illustrates all the above mensioned phenomena.
In the Fig.\ref{hyper-nonoise}, we consider the Hyperchaos system of R\"{o}ssler\cite{rossler}:
\begin{eqnarray*}
\dot{x}&=&-y-z,~~~~~~~~~~ \dot{y}=x+0.25y+w, \\
\dot{z}&=&3+xz,~~~~~~~~~~ \dot{w}=-0.5z+0.05w,
\end{eqnarray*}
with the sampling time $T_{s}=0.1$ and the initial condition $(-20,0,0,15)$.
We acquire $50000$ points with $v(k)=x(kT_{s})$ and then for calculating $D(N,r,d,T_{d})$, 5000 points are picked up from the data set, satisfied by
\begin{equation}
j=\frac{[N_{T}-(d-1)T_{d}-1]}{N-1}
\end{equation}
in Eq. (\ref{emb-vec}). It is important to evenly pick up the data points because the full time series has to be reflected.
As the algorithm to integrate the differential equations, we adopt the fourth order Runge-Kutta method\cite{william}.

\begin{figure}
\epsfig{figure=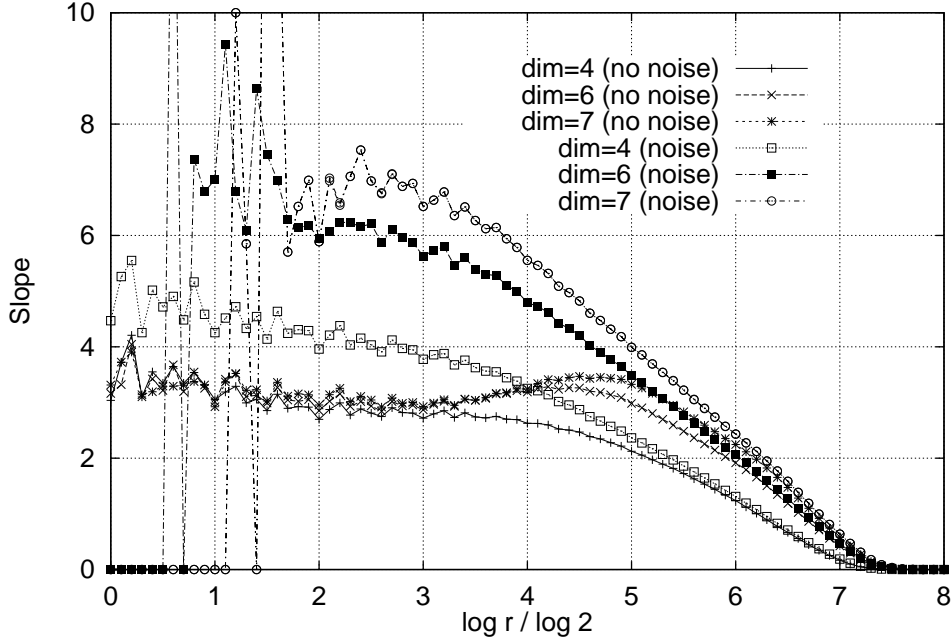}
\caption{Slope$=d[\log_{2}C_{d}(r)]/d\log_{2}(r)$ vs $\log_{2}(r)$ for $50000$ points generated by the Hyperchaos system of R\"{o}ssler and Slope vs $\log_{2}(r)$ for $50000$ points generated by the Hyperchaos system which has been contaminated with the gaussian white noise of the mean $0$  and the variance $25$. $C_{d}$ means the correlation integral with the embedding dimension $d$. While the plot of slope vs $\log_{2}(r)$ for the noise-free data tells that the corrlation dimension is about $3$, the plot for the noisy data is completely the one for a structureless random noisy data. The delay time $T_{d}$ is $30$.}  \label{correlation-integral}
\end{figure}

\begin{figure}
\epsfig{figure=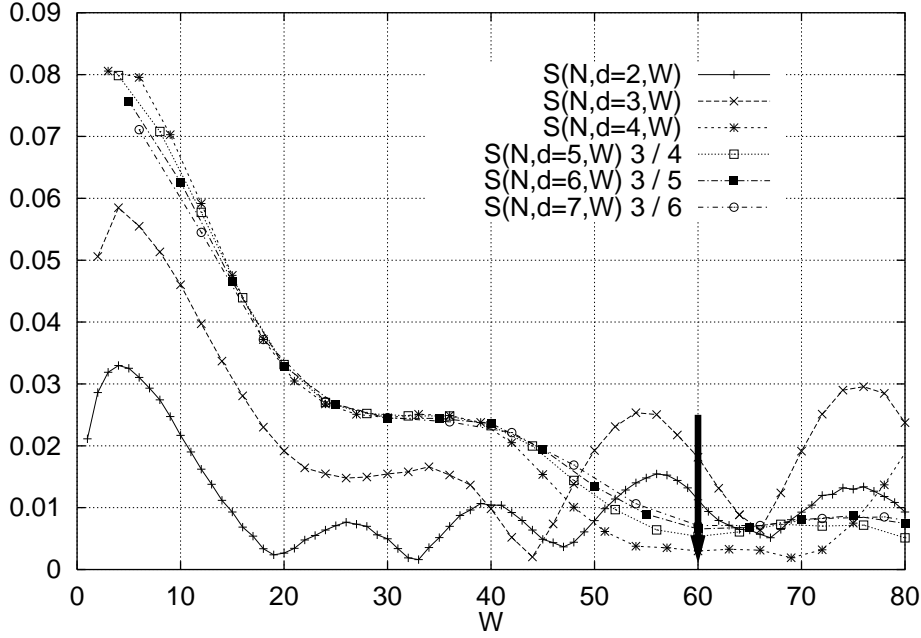}
\caption{$S(N,d,W)$ vs window length $W$ for $N=49500$ among $50000$ points generated by the Hyperchaos system of R\"{o}ssler which has been contaminated with the gaussian white noise of the mean $0$  and the variance $25$. For the sake of comparison, each $S(N,d,W)$ was appropriately scaled. Despite of the presence of the measurement noise, everything is almost the same as Fig. \ref{hyper-nonoise}. We can say that $d_{M}=4$ and $T_{d}^{d=4} \approx 20$. $\Delta r = 2$, $\Delta T_{d} = T_{s} = 0.1$.}  \label{hyper-noise}
\end{figure}

In this paper, we want to place emphasis on the robustness of this method against measurement noise.
A time series $V(k)$ contaminated by a measurement noise is represented as follows:
\begin{equation}
V(k) = v(k) + \lambda(k),  \label{noise-eq}
\end{equation}
where $v(k)$ is a noise-free time series and $\lambda(k)$ a measurement noise.
The effect of measurement noise $\lambda(k)$ in Eq.(\ref{noise-eq}) in the reconstruction procedure results only in the thickness of the reconstructed trajectory.
However, in the variation with respect to the delay time $T_{d}$ (Eq.(\ref{sum-diff})), the thickness due to noise is canceled out because it has nothing to do with the delay time $T_{d}$.
To show the robustness against noise, we make a time series $V(k)=x(k)+\lambda(k)$, where $x(k)$ is just the time series for Fig.\ref{hyper-nonoise} and $\lambda(k)$ is a gaussian white noise with the mean `0' and the variance `25'. This time series $V(k)$ is too noisy to be discriminated from a structureless noisy time series by means of the correlation integral(Fig.\ref{correlation-integral}).
The result is shown in Fig.\ref{hyper-noise}. One should notice that the trend of variation is totally similar to the one of the measurement noise-free time series(Fig.\ref{hyper-nonoise}).

\begin{figure}
\epsfig{figure=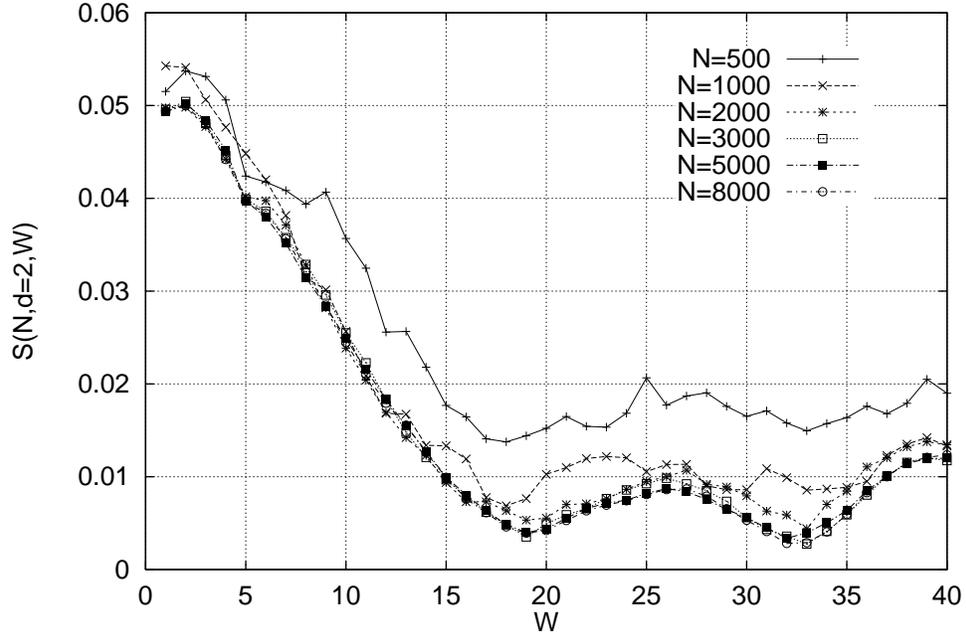}
\caption{$S(N,d=2,W)$ vs window length $W$ for various values of $N$, evenly picked up among $50000$ points generated by the Hyperchaos system of R\"{o}ssler with $\Delta r=2$, $\Delta T_{d}=T_{s}=0.1$. The curves for $S(N,d=2,W)$ are saturated at about $N=3000$.}  \label{pickup-number}
\end{figure}

In evaluating $D(N,r,d,T_{d})$, it takes exponentially increasing time according to the number $N$ of picked-up points.
Therefore, if one wants to pay a lower cost for the $D(N,r,d,T_{d})$, it is important to pick up points as few as possible, which depends closely on what values of $\Delta T_{d}$ and $\Delta r$ are set.
Because the purpose of calculation of $D(N,r,d,T_{d})$ is to infer globally the spatial distribution of a reconstructed attractor, not to examine the fine structure, it is sufficient to set $\Delta r \approx R/50$, where $R$ is the largest length among the distances between any two points in the reconstructed attractor.
On the other hand, using the power spectrum of time series, the $\Delta T_{d}$ can be properly chosen. This is because in $d=2$, the first minimum point $T_{d}^{d=2}$ is, on the average, related to the temporal fluctuation of time series.
As a guide line for the choice of $\Delta T_{d}$, we advise $1/(4f_{h}\Delta T_{d}) \approx 10$, where $f_{h}$ means the highest frequency in the power spectrum of the time series. The contribution of background noise has to not be considered.
This condition corresponds to $W^{d=2} \approx 10\Delta T_{d}$. The sampling time $T_{s}$ has to be $\leq \Delta T_{d}$.
With these crude guidance, we chose the values for $\Delta T_{d}$ and $\Delta r$ in Fig.\ref{hyper-nonoise} and Fig.\ref{hyper-noise}.
As the last and the most important step, with the roughly chosen $\Delta T_{d}$ and $\Delta r$, one has to find the proper number $N$ of picked-up points. This is easily done by finding when $D(N,r,d=2,T_{d})$ is saturated. This step is shown in Fig.\ref{pickup-number} which indicates us that the proper $N$ for Fig.\ref{hyper-nonoise} is about 3000, even though we use $N=5000$ in Fig.\ref{hyper-nonoise}.


\newpage

\newpage

\newpage

\end{document}